\documentclass[aps,pra,twocolumn,floatfix,showpacs]{revtex4}
\usepackage{amsmath}
\usepackage{amsfonts}
\usepackage{amssymb}
\usepackage{graphicx}
\usepackage{color}

\begin{document}

\title{Fano resonances and their control in optomechanics}
\author{Kenan Qu and G. S. Agarwal}
\affiliation{Department of Physics, Oklahoma State University, Stillwater, Oklahoma, 74078, USA}
\date{\today}

\begin{abstract}
The Fano line profiles, originally discovered in the context of photoionization, have been found to occur in a large class of systems like resonators, metamaterials, plasmonics. We demonstrate the existence of such resonances in cavity optomechanics by identifying the interfering contributions to the fields generated at antiStokes and Stokes frequencies. Unlike the atomic systems, the optomechanical systems provide great flexibility as the width of the resonance is controlled by the coupling field. We further show how the double cavities coupled by a single optomechanical mirror can lead to the splitting of the Fano resonance and how the second cavity can be used to tune the Fano resonances. The Fano resonances are quite sensitive to the decay parameters associated with cavities and the mechanical mirror. Such resonances can be studied by both pump probe experiments as well as the spectrum of the quantum fluctuations in the output fields.
\end{abstract}
\maketitle

\section{Introduction}
The Fano line shapes~\cite{Fano} have played an important role in our understanding of the photoelectron spectra~\cite{Bransden} in atomic physics and more recently in the field of plasmonics~\cite{plasmonics}. In atomic systems the interference minimum in the Fano line shape has been extensively used in extracting information on the relative strengths of the different decay channels which are responsible for interference effects. The Fano interference has been at the forefront in producing lasing without population inversion~\cite{LWI}. The Fano line shapes have been considered as a probe of decoherence~\cite{note1}. The Fano interference is leading to other remarkable  possibilities like the improvement in the efficiency of the heat engines~\cite{heatengine}. In plasmonics, interference effects play a very important role as photons can travel along different interfering paths and thus the Fano line shapes become quite common. The asymmetry of the Fano line shape can be continuously tuned by controlling the rate of saturation of two interference paths~\cite{fanoRMP}. Such line shapes are also used in obtaining information on the interaction between nano particles with light~\cite{plasmonics}. Interestingly enough the Fano line shapes can be understood rather easily by using a two coupled oscillator model in which the two oscillators have quite different lifetimes~\cite{AJP}. Thus the lifetimes of the modes are important in the realization of the Fano line shapes. In fact, plasmonic systems and metamaterial structures are designed keeping the lifetime issue in mind~\cite{plasmonics}. It turns out that the lifetime conditions are well satisfied for the optomechanical systems (OMS) since the damping of the mirror is much smaller than the cavity damping.

In the last few years the OMS have shown the possibility of a variety of interference effects~\cite{OMIT,Painter}. Clearly such a system, under certain condition, can exhibit Fano line shape~\cite{Fano-OMS1,Fano-OMS2} and in fact \cite{Painter} reports the observation of the Fano line shape. In this paper, we analyze the physics and mathematics behind Fano resonances in optomechanics and report the possibility of double Fano resonance in coupled OMS. The double Fano resonances which are interesting in their own right can provide information on a variety of optomechanical couplings. The Fano minima can also be exploited for the successful transfer of states and transduction of photons. The double cavity set up can lead to the tunability of the Fano resonances, which in fact has been an important issue in other contexts~\cite{gu}. The double-cavity OMS are not the only ones for the observation of double Fano resonances, for example, we can use electromechanical systems with several superconducting capacitors~\cite{qlimit}.

The outline of this paper is as follows: In Sec.~\ref{analyze}, we recall the Fano line shapes and physically explain how these can arise in OMS. In Sec.~\ref{model}, we introduce the model for OMS and derive the line profiles under steady state conditions. In Sec.~\ref{single}, we examine the Fano resonance in a single cavity OMS and derive its resonance frequency center as well as its width. We also quantify the Fano shape by using the Fano asymmetry parameter. In Sec.~\ref{double}, we consider the double-cavity OMS and show the existence of the double Fano lineshape in different coupling regimes. We explain how we can tune Fano resonances by changing the coupling fields. In Sec.~\ref{quantum}, we calculate the quantum fluctuations of the fields in OMS, and demonstrate the existence of Fano resonances in the spectrum of quantum fluctuations.

\section{Fano resonances from interfering paths in cavity optomechanics} \label{analyze}
In the classic work~\cite{Fano}, Fano considered photoionization process when a weakly bound state $|a\rangle$ lies in the continuum $|E\rangle$. The weakly bound state has a finite life time due to its coupling with the continuum. Thus, in the simplest case, there are two transition amplitudes leading to photoionization---one involves a direct transition to the continuum and the other involves transition via the autoionizing state to the continuum. These two transition amplitudes interfere leading to the famous Fano formula for the probability $p$ for the photoionization
\begin{equation}\label{0}
    p(E)=\frac{(\epsilon+q)^2}{\epsilon^2+1}, \qquad \epsilon=\frac{2(E-E_a)}{\Gamma},
\end{equation}
where $q$ is called the Fano asymmetry parameter and $\Gamma$ is the width of the state $|a\rangle$. The Fano minimum occurs at $\epsilon=-q$. The parameter $q$ depends on the relative strengths of the independent transitions to the state $|a\rangle$ and $|E\rangle$. If $q$ is large, interference disappears.

\begin{figure}[phtb]
 \includegraphics[width=0.4\textwidth]{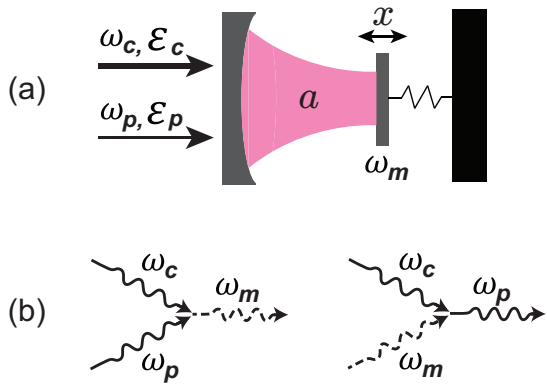}
 \caption{\label{Fig1}{(Color online) (a) Schematic of OMS and (b) photon-phonon interaction.}}
\end{figure}

Clearly, to produce Fano line shapes, we need two paths which interfere. Let us see, in terms of the physical processes, the possible physical mechanisms which lead to the building up of the cavity field in an OMS as shown in Fig.~\ref{Fig1}a.
Here the field $\mathcal{E}_p$ with frequency $\omega_p$ is close to the cavity frequency $\omega$ and the field $\mathcal{E}_c$ is red detuned from the cavity resonance by an amount close to the oscillating frequency of the mirror. If the mirror position $x$ were fixed, the cavity field at $\omega_p$ will have the form $\mathcal{E}_p/[\kappa+i(\omega-\omega_p)]$, where $2\kappa$ is the rate at which the photons leak out of the left mirror. If the mirror is active, due to the radiation pressure interaction, we have the possibility of Stokes and antiStokes processes. The fields $\mathcal{E}_c$ and $\mathcal{E}_p$ can produce a coherent photon $x(\neq0)$. Once the coherent phonon is produced it can combine with $\mathcal{E}_c$ to produce the cavity field at $\omega_p$. In the case of a moving mirror, the two successive nonlinear frequency conversion processes between two light fields and a phonon as illustrated in Fig.~\ref{Fig1}b would produce the cavity field. Therefore, in OMS, we have two coherent processes leading to the building up of the cavity field. These are: (a) direct building up due to the application of $\mathcal{E}_p$; and (b) building up due to the nonlinear processes in Fig.~\ref{Fig1}b. These two processes interfere leading to the Fano resonances in OMS. The precise mathematical treatment is given in Sec.~IV, where we introduce the analog of the Fano asymmetry parameter $q$.

In atomic physics, Fano profile have been extensively studied. Many generalizations to several resonances in continuum or interactions with different continua exist. The different continua can be either radiative or ionizing. Radiative continuum tends to make the value of the minimum nonzero~\cite{note1}.

\section{Basic Model} \label{model}
\begin{figure}[phtb]
 \includegraphics[width=0.35\textwidth]{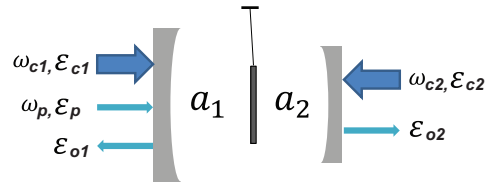}
 \caption{\label{Fig2}{(Color online) Schematic double-cavity OMS.}}
\end{figure}

We study the system in Fig.~\ref{Fig2}, in which a mechanical resonator, coated with perfect reflecting films on both sides, is coupled to two cavities.
The mechanical resonator is modeled as a harmonic oscillator with mass $m$, frequency $\omega_m$ and momentum decay rate $\gamma_m$. For each cavity, we denote its field by $a_i$, frequency $\omega_i$ and decay rate $\kappa_i$, $i=1,2$. The field annihilation and creation operators satisfy the commutation relation $[a_i,a^\dag_j]=\delta_{ij}$. The probe laser with frequency $\omega_p$, is sent into cavity $1$. Other possible realizations would be using ``membrane in the middle'' setup~\cite{membrane}. The two cavities are coupled only via the oscillations of the mechanical mirror which are produced by the applied strong laser fields $\mathcal{E}_{ci}$'s. Further the two cavities can be in different frequency regimes. We introduce normalized coordinates $Q=\sqrt{m\omega_m/\hbar}x$ and $P=\sqrt{1/(m\hbar\omega_m)}p$ for the mechanical oscillator with commutation relations $[Q,P]=[x,p]/\hbar=i$. The Hamiltonian for this system is given by
\begin{align}\label{1}
  H &= H_1 + H_2 + H_m + H_\text{diss}, \nonumber \\
  H_1 &= \hbar(\omega_1-\omega_{c1})a_1^\dag a_1 - \hbar g_1 a_1^\dag a_1 Q + i\hbar\mathcal{E}_{c1}(a_1^\dag-a_1) \nonumber \\
   &\qquad + i\hbar(\mathcal{E}_p a_1^\dag \mathrm{e}^{-i\delta t} - \mathcal{E}_p^* a_1 \mathrm{e}^{i\delta t}) \nonumber \\
  H_2 &= \hbar(\omega_2-\omega_{c2})a_2^\dag a_2 + \hbar g_2 a_2^\dag a_2 Q + i\hbar\mathcal{E}_{c2}(a_2^\dag-a_2) \nonumber \\
  H_m &= \frac12\hbar\omega_m(P^2+Q^2),
\end{align}
where $\delta=\omega_p-\omega_{c1}$ is the detuning between the probe field and the coupling field in cavity $1$; and coupling coefficients are defined by $g_i=(\omega_{i}/L_i)\sqrt{\hbar/(m\omega_m)}$ with $L_i$ being the length of the $i$th cavity. The $\mathcal{E}$ terms in (\ref{1}) denote the coupling of the cavity field to the applied laser fields. The coupling and probe fields are related to the power of the applied laser fields via  $\mathcal{E}_{ci} = \sqrt{2\kappa_iP_i/(\hbar\omega_i)}$, and $i=1,2$ and $\mathcal{E}_p = \sqrt{2\kappa_1P_{p}/(\hbar\omega_p)}$, respectively. In Eq.~(\ref{1}), all the dissipative interactions are denoted by $H_\text{diss}$. These include the leakage of photons from both cavities and the Brownian motion of the mirror. The Hamiltonian (\ref{1}) has been written by working in a picture so the very fast frequencies $\omega_{ci}$'s are removed. This results in detuning terms like $(\omega_i-\omega_{ci})a_i^\dag a_i$.

Using Eq. (\ref{1}) we can derive the quantum Langevin equations for the operators $Q$, $P$, $a_i$ and $a_i^\dag$. However, for the purpose of this paper, we will work with semiclassical equations so that all operator expectation values are replaced by their mean values. Thus in the rest of the paper, all the quantities $Q$, $P$, $a_i$ and $a_i^\dag$ will be numbers. The equations of motion for $Q$, $P$, $a_i$ and $a_i^\dag$  are found to be
\begin{equation}\label{2}
\begin{aligned}
		\dot{Q} & =\omega_{m}P, \\
		\dot{P} & =(g_{1}a_{1}^{\dagger}a_{1}-g_{2}a_{2}^{\dagger}a_{2})-\omega_{m}Q-\gamma_{m}P, \\
		\dot{a}_{1} & = -i(\omega_1-\omega_{c1}-g_1Q)a_{1} -\kappa_{1}a_{1} +\mathcal{E}_{c1} +\mathcal{E}_{p}\mathrm{e^{-i\delta t}}, 	\\
		\dot{a}_{2} & = -i(\omega_2-\omega_{c2}+g_2Q)a_{2}-\kappa_{2}a_{2}+\mathcal{E}_{c2}.
\end{aligned}
\end{equation}
Eqs. (\ref{2}) involve periodically oscillating terms hence in the long time limit, any of the fields and the mechanical coordinates will have a solution of the form $A=\sum_{n=-\infty}^{+\infty} \mathrm{e}^{-in\delta t}A_n$. The $A_n$'s can be obtained by the Floquet analysis. We assume that the probe is much weaker than the coupling field, then $A_n$'s can be obtained perturbatively. In particular, we find the steady state results to first order in $|\mathcal{E}_{p}/\mathcal{E}_{ci}|$
\begin{align}
		a_{10} &= \frac{\mathcal{E}_{c1}}{\kappa_1+i\Delta_1}, \qquad a_{20} = \frac{\mathcal{E}_{c2}}{\kappa_2+i\Delta_2}, \label{7}\\
        Q_0 &= \frac{1}{\omega_m}(g_1|a_{10}|^2-g_2|a_{20}|^2), \label{8}\\
		Q_+ &= -\frac{1}{d(\delta)}\frac{g_1a_{10}^*\mathcal{E}_{p}}{(\kappa_1+i\Delta_1 -i\delta)}, \label{9}\\
		d(\delta) &= \sum_{i=1,2} \frac{2\Delta_ig_i^2|a_{i0}|^2}{(\kappa_i-i\delta)^2 + \Delta_i^2} - \frac{\omega_m^2-\delta^2-i\delta\gamma_m}{\omega_m}, \label{10} \\
		a_{1+} &= \frac{ig_1a_{10}}{(\kappa_1+i\Delta_1-i\delta)}Q_+ + \frac{\mathcal{E}_p}{(\kappa_1+i\Delta_1-i\delta)}, \label{11} \\
		a_{1-} &= \frac{ig_1a_{10}}{(\kappa_1+i\Delta_1+i\delta)}Q_+^*, \label{12} \\
    	a_{2+} &=\frac{-ig_2a_{20}}{(\kappa_2+i\Delta_2-i\delta)}Q_+,		 \label{13}\\
    	a_{2-} &=\frac{-ig_2a_{20}}{(\kappa_2+i\Delta_2+i\delta)}Q_+^*,		 \label{14}
\end{align}
where $\Delta_1=\omega_1-\omega_{c1}-g_1Q_0$ and $\Delta_2=\omega_2-\omega_{c2}+g_2Q_0$ are the detunings of the coupling lasers to the effective cavity frequencies. $Q_0$, which is typically small, denotes the displacement of the mechanical resonator under radiation pressure. The cavity field $a_{i0}$ is the field in the $i$th cavity at the frequency of the coupling laser. The coupling efficients are enhanced by the cavity photon number hence we define $G_i=g_i|a_{i0}|/\sqrt2$. The fields $a_{i\pm}$'s are the antiStokes and Stokes fields in the $i$'th cavity. The output fields from the two cavities are given by
\begin{align}
    \mathcal{E}_{1out} &= 2\kappa_1(a_{10}\mathrm{e}^{-i\omega_{c1} t} + a_{1+}\mathrm{e}^{-i(\omega_{c1}+\delta) t} + a_{1-}\mathrm{e}^{-i(\omega_{c1}-\delta) t}) \nonumber \\
     &\quad - \mathcal{E}_p\mathrm{e}^{-i\omega_p t} \label{15}\\
    \mathcal{E}_{2out} &= 2\kappa_2(a_{20}\mathrm{e}^{-i\omega_{c2} t} + a_{2+}\mathrm{e}^{-i(\omega_{c2}+\delta) t} + a_{2-}\mathrm{e}^{-i(\omega_{c2}-\delta) t}) \label{16}
\end{align}

We now concentrate on the output fields from the cavity $1$. The output fields in the form of Eq.~(\ref{15}) contain components at three different frequencies: the coupling frequency $\omega_{c1}$; the antiStokes frequency, which is also the probe frequency, $\omega_{c1}+\delta=\omega_p$; and the Stokes frequency $\omega_{c1}-\delta$. Among these three components, we are most interested in the generated antiStokes and Stokes sidebands and we display them as the normalized quantities defined by $\mathcal{E}_{ias} = 2\kappa_ia_{i+}/\mathcal{E}_p$ and $\mathcal{E}_{is} = 2\kappa_ia_{i-}/\mathcal{E}_p$. The actual normalized output field at the antiStokes frequency from the cavity $1$ is given by $(\mathcal{E}_{1as}-1)$, cf. Eq.~(\ref{15}). The antiStokes field would be resonantly enhanced in the vicinity of the cavity frequency $\omega_1$, when both the coupling fields are tuned by an amount close to the mechanical frequency below their corresponding cavity frequency, i.e. $\Delta_1\sim\omega_m$. We work in the regime with cooperativity $C_i=G_i^2/\kappa_i\gamma_m>1$ in which the OMS is strongly coupled, then the antiStokes and Stokes fields in cavity $1$ are given by
\begin{align}
    \mathcal{E}_{1as} &= \frac{2\kappa_1}{[\kappa_1-i(\delta-\Delta_1)] + \frac{G_1^2}{[\frac{\gamma_m}{2}-i(\delta-\omega_m)] + \frac{G_2^2}{[\kappa_2-i(\delta-\Delta_2)]}}}, \label{17} \\
    \mathcal{E}_{1s}  &= \frac{-i\kappa_1/\omega_m}{1 + \frac{[\kappa_1+i(\delta-\Delta_1)]}{G_1^2}\left\{ [\frac{\gamma_m}{2}+i(\delta-\omega_m)] + \frac{G_2^2}{\kappa_2+i(\delta-\Delta_2)]}\right\} }. \label{18}
\end{align}
Similarly, the antiStokes and Stokes fields in cavity $2$ are found to be
\begin{align}
    \mathcal{E}_{2as} &= \frac{-2\kappa_2 \frac{G_1}{[\kappa_1-i(\delta-\Delta_1)]} \frac{G_2}{[\kappa_2-i(\delta-\Delta_2)]}}{\frac{G_1^2}{[\kappa_1-i(\delta-\Delta_1)]} + \frac{G_2^2}{[\kappa_2-i(\delta-\Delta_2)]} + [\frac{\gamma_m}{2}-i(\delta-\omega_m)]}, \label{19} \\
    \mathcal{E}_{2s}  &= \frac{\kappa_2\frac{G_2}{\omega_m} \frac{G_1}{[\kappa_1-i(\delta-\Delta_1)]}}{\frac{G_1^2}{[\kappa_1+i(\delta-\Delta_1)]} + \frac{G_2^2}{[\kappa_2+i(\delta-\Delta_2)]} + [\frac{\gamma_m}{2}+i(\delta-\omega_m)]}. \label{20}
\end{align}

\section{The Fano Resonance in the output fields} \label{single}
\begin{figure}[phtb]
 \includegraphics[width=0.4\textwidth]{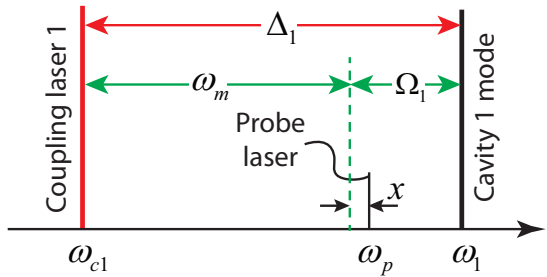}
 \caption{\label{Fig3}{(Color online) Schematic illustration of frequencies used in obtaining Fano lineshapes. Fano asymmetry parameter $q$ is defined in terms of detuning $q=-\Omega_1/\kappa_1$. The effective damping is defined by $\Gamma_1=G_1^2/[\kappa_1(1+q^2)]$.}}
\end{figure}

\begin{figure}[phtb]
 \includegraphics[width=0.45\textwidth]{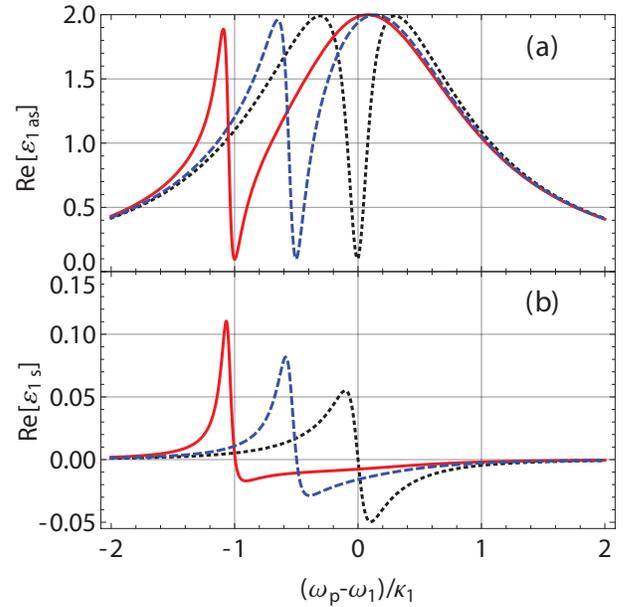}
 \caption{\label{Fig4}{(Color online) The antiStokes field $\mathcal{E}_{as}$ (a) and the Stokes field $\mathcal{E}_{s}$ (b) as a function of frequency of the probe laser input $\omega_p$ for the OMS in Fig.~\ref{Fig1}a. The black dotted, blue dashed and red solid curves are corresponding to $\Omega_1=0$, $\Omega_1=0.5\kappa_1$ and $\Omega_1=\kappa_1$, respectively.}}
\end{figure}

We examine now Fano resonances in the output fields. We have four different fields $\mathcal{E}_{ias}$, $\mathcal{E}_{is}$, and $i=1,2$. We first examine $\mathcal{E}_{1as}$ if $G_2=0$, i.e. the case of OMS in Fig.~\ref{Fig1}. For this system, the antiStokes field is
\begin{equation}\label{21}
    \mathcal{E}_{1as} = \frac{2\kappa_1[\frac{\gamma_m}{2}-i(\omega_p-\omega_{c1}-\omega_m)]}{[\kappa_1-i(\omega_p-\omega_1)][\frac{\gamma_m}{2}-i(\omega_p-\omega_{c1}-\omega_m)] + G_1^2}.
\end{equation}
Typically the mechanical damping is much smaller than the cavity damping, $\gamma_m\ll\kappa_1$, and we work in the resolved sideband limit, $\omega_m\gg\kappa_1$. We expect two resonances at $\omega_p\approx\omega_1$, and at $\omega_p\approx\omega_{c1}+\omega_m=\omega_1+(\omega_m-\omega_1+\omega_{c1})=\omega_1-\Omega_1$. In order to keep these two resonances distinct, we keep $\omega_1\neq\omega_{c1}+\omega_m$, i.e. $\Omega_1\neq0$. For clarity, we show the relations between different frequencies in Fig.~\ref{Fig3}.
The resonance at $\omega_p=\omega_{c1}+\omega_m$ would be very narrow since $\gamma_m\ll\kappa_1$. The frequency offset factor $\Omega_1$ plays important role in the production of the Fano line shapes. Physically it means that the antiStokes process is not resonant with the cavity frequency. We examine the structure of $\mathcal{E}_{as1}$ near the resonance $\omega_p=\omega_{c1}+\omega_m=\omega_1-\Omega_1$ for a fixed value of $\Omega_1$, and we define $x=\omega_p-\omega_1+\Omega_1$. In the vicinity of this resonance, $x\sim0$ and Eq.~(\ref{21}) can be approximated to
\begin{align}\label{22}
    \mathcal{E}_{1as} &\approx \frac{ 2\kappa_1(\gamma_m/2-ix)}{(\kappa_1+i\Omega_1)(\gamma_m/2-ix) + G_1^2} \nonumber \\
    &\approx \frac{2\kappa_1}{\kappa_1+ i \Omega_1} \cdot \frac{x }{x + \frac{i G_1^2}{\kappa_1+i\Omega_1} } \nonumber \\
    &= \frac{2}{1+ i\frac{\Omega_1}{\kappa_1}} \cdot \frac{\frac{\kappa_1^2+\Omega_1^2}{\kappa_1G_1^2}x }{ \frac{\kappa_1^2+\Omega_1^2}{\kappa_1G_1^2}x+\frac{\Omega_1}{\kappa_1}+i},
\end{align}
and hence
\begin{equation}\label{23}
    Re[\mathcal{E}_{1as}] = \frac{2}{1+q^2}\cdot\frac{(\bar{x}+q)^2}{\bar{x}^2+1},
\end{equation}
where $\bar{x}=x/\Gamma-q$, $\Gamma\cong\frac{\kappa_1G_1^2}{\Omega_1^2+\kappa_1^2}$, and $q=-\Omega_1/\kappa_1$. The profile (\ref{23}) has exactly the same form as the classic profile of Fano resonance with maximum at $\bar{x}=1/q$ and zero at $\bar{x}=-q$. The asymmetry parameter $q$ is related to the frequency offset $\Omega_1$. Keep in mind that this is derived in the vicinity of $x\sim0$, i.e. $\omega_p\simeq\omega_1-\Omega_1$. In order to see explicitly the nature of the output fields, we use the following set of experimentally realizable parameters $\omega_m=2\pi\times10$MHz, $\gamma_m=2\pi\times0.01$MHz, $\kappa_1=2\pi\times1$MHz, and $G_1=2\pi\times0.3$MHz. We display the full profile of the output fields in Fig.~\ref{Fig4}a as a function of $(\omega_p-\omega_1)/\kappa_1$  for a single cavity OMS.
It shows the narrow Fano profile as well as the relatively broad resonance near $\omega_p\sim\omega_1$. For detuning $\Delta_1=\omega_m$ (dotted curve), we obtain the standard EIT profiles~\cite{OMIT,Painter}. As we increase the detuning, the Fano resonance shifts away from the cavity resonance frequency and becomes asymmetric. Each of these Fano lineshapes has a zero point exactly at the frequency $\bar{x}=-q$ or equivalently $\omega_p-\omega_1=-\Omega_1$. Our approximation formula (\ref{23}) and the numerical curves obtained directly from (\ref{17}) agree well.

Safavi-Naeini et al~\cite{Painter} have observed such profiles for a broad range of $q$ values. What we have demonstrated in this section is how Fano line shapes can arise in OMS under the condition $\gamma_m\ll\kappa_1$. When $\gamma_m$ starts increasing, the character of the line shape starts changing in a manner similar to changes in the Fano line profiles when the radiative effects are included.

It is also noteworthy to study the Stokes sideband generated by the coupling laser and the mechanical oscillator, although it is suppressed because it is an off-resonantly process. In Fig.~\ref{Fig4}b, we plot the Stokes sideband. The line shape is asymmetric though a good signature of interference is missing. This is because Fano resonance requires two coherent routes for building up the cavity field, which can interfere with each other, whereas the only route producing Stokes sideband is via the combination of coupling field and the mechanical phonons.

\section{Double Fano Resonances in Cavity Optomechanics} \label{double}
Recently double cavity configurations have attracted a lot of attention because of their wide applicability in photon switching~\cite{PPT,KQ}, state transfer~\cite{Tian,Clerk} and transduction of photons~\cite{Vitali}. We discuss yet another possibility, making use of double cavities to tune the Fano resonances. In this section, we will show how we can change and control the Fano resonance by adding a second cavity in OMS. When the coupling fields exist in both cavities, the denominator of Eq.~(\ref{17}) becomes cubic and hence the number of roots increases from two to three. This is because we have three coupled systems: two cavity modes and one mechanical mode. At the same time, the numerator in~(\ref{17}) becomes a quadratic function of $x$ suggesting the possibility of two different minima in the output fields. Therefore, a single Fano resonance goes over to a double Fano resonance.

Next we examine the quantitative features of the double Fano resonance in OMS. The parameter space is large and therefore we begin by fixing the detuning in cavity $1$ as $\Delta_1=\omega_m+\kappa_1$ so that its Fano asymmetry parameter $q=-1$, and we let the detuning of cavity $2$ arbitrary such that $\Delta_2=\omega_m+\Omega_2$. In the vicinity of $x=\omega_p-\omega_1+\kappa_1\sim0$, the roots of the numerator in Eq.~(\ref{17}) determine the existence of the Fano minima. We first discuss the case when $G_2^2\gg\Omega_2^2$. Then to first order in dampings, the roots are
  \begin{equation}\label{32}
    x_\pm \simeq \pm G_2+\frac{\Omega_2}{2} - i\frac{\kappa_2+\gamma_m/2}{2},
  \end{equation}
leading to the splitting of the Fano resonances into two. The power of the coupling field in cavity $2$ determines their frequency splitting.
\begin{figure}[phtb]
 \includegraphics[width=0.45\textwidth]{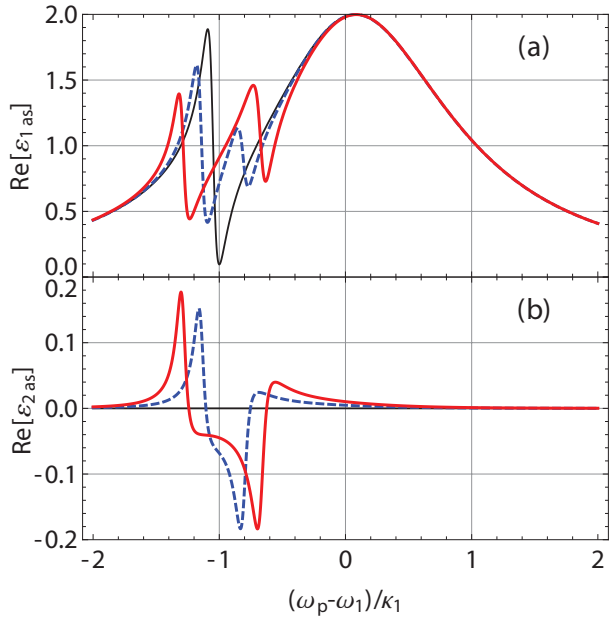}
 \caption{\label{Fig5}{(Color online) The antiStokes fields $\mathcal{E}_{as}$ in cavity $1$ (a) and in cavity $2$ (b) as a function of frequency of the probe laser input in a double-cavity OMS. The thin black, blue dashed and red solid curves are corresponding to different coupling strength of cavity $2$ that $G_2/\kappa_1=0$, $0.15$ and $0.3$, respectively. We set $\Omega_2=0.1\kappa_1$ and $\kappa_2=0.05\kappa_1$.}}
\end{figure}
In Fig.~\ref{Fig5}a, we explicitly show the splitting of the Fano resonance in the double-cavity OMS using the same parameters for cavity $1$ and $\kappa_2=0.05\kappa_1$, $\Omega_2=0.1\kappa_1$ for cavity $2$ with different coupling strengths. In Fig.~\ref{Fig5}a, the thin curve shows a single Fano resonance when the coupling field in cavity $2$ is absent. As we increase the coupling field in cavity $2$, the Fano resonance splits and the splitting increases linearly as we increase the coupling power. Apart from the splitting, their resonance frequency center is shifted by an amount $\Omega_2/2$. The frequency splittings of the two Fano resonance are $0.3\kappa_1$ and $0.6\kappa_1$, which respectively equals to $2G_2$. The frequency splitting is independent of the detuning of cavity $2$, as long as it is close to $\omega_m$. Therefore, one can always obtain the coupling strength, as well as the coupling power, by measuring the double Fano resonances. In the figure, the minimum values of the double Fano resonances do not go to zero due to the finite values of $\kappa_2$ and $\gamma_m$. In an OMS with lower $\kappa_2$, we should be able to obtain a lower minimum and a higher maximum in the double Fano resonances. This is reminiscent of the result in the context of photoionization in which the value of the minimum depends on the radiative effects~\cite{note1}.

In Fig.~\ref{Fig5}b, we plot the antiStokes field in cavity $2$ in response to the probe laser input in cavity $1$. We see asymmetric peaks generated around the frequency of the Fano resonances in $\mathcal{E}_{1as}$ and their widths are similar to the corresponding Fano resonances. Both the peaks heights and peaks splitting increase with the increasing of the coupling power. Physically, this can be interpreted as the probe energy in cavity $1$ is transferred to cavity $2$ via the mechanical resonator. The antiStokes field in cavity $2$ shows anti-symmetric split Fano resonances.

\begin{figure}[phtb]
 \includegraphics[width=0.45\textwidth]{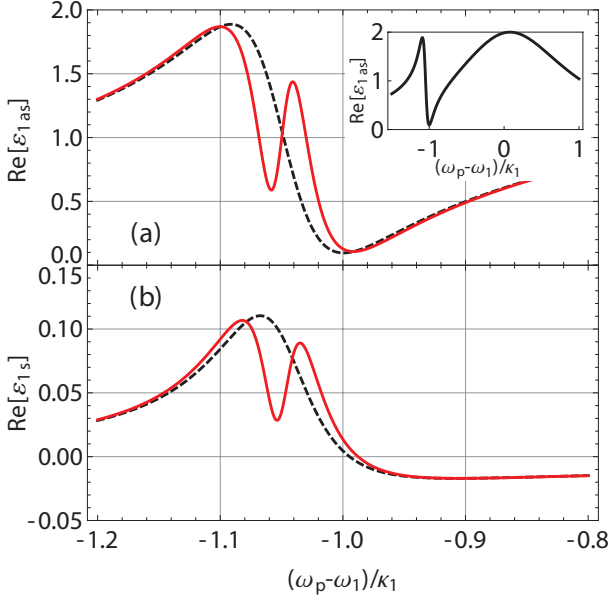}
 \caption{\label{Fig6}{(Color online) The antiStokes field (a) and Stokes field (b) in cavity $1$ as a function of frequency of the probe laser input in a double-cavity OMS. The solid curves: double-cavity OMS with $G_2=0.02\kappa_1$ shows the emergence of a the second Fano resonance, compared to the dashed curve: single cavity OMS. The inset in (a) shows the full profile in a large scale. We set $\Omega_2=-0.05\kappa_1$ and $\kappa_2=0.5\gamma_m=0.005\kappa_1$. }}
\end{figure}

The characteristics of the double Fano resonances are different in the weak coupling limit. When $G_2^2 \ll \Omega_2^2$, the roots of the numerator in Eq.~(\ref{17}) determining the Fano minima are
  \begin{equation}\label{32}
  \begin{aligned}
    x_+ &\simeq -\frac{G_2^2}{\Omega_2} - i\kappa_2\frac{G_2^2}{\Omega_2^2} - i\frac{\gamma_m}{2}(1-\frac{G_2^2}{\Omega_2^2}) \\
    x_- &\simeq \Omega_2 - i\kappa_2(1-\frac{G_2^2}{\Omega_2^2}) - i\frac{\gamma_m}{2}\frac{G_2^2}{\Omega_2^2} .
  \end{aligned}
  \end{equation}
The root $x_+$ indicates a frequency shift of the Fano resonance with an amount $-G_2^2/\Omega_2$, and the root $x_-$ implies the emergence of a new Fano resonance around $x\sim\Omega_2$ besides the original Fano resonance around $x\sim0$. In Fig.~\ref{Fig6}, we illustrate both the antiStokes and Stokes field in cavity $1$ using the following parameters $\kappa_2=0.5\gamma_m=0.005\kappa_1$, $\Omega_2=-5\gamma_m=-0.05\kappa_1$, and $G_2=0.02\kappa_1$, (compared with $G_2=0$ for the single cavity case as dashed curves) and parameters for cavity $1$ are identical to Fig.~\ref{Fig4}.
Using these parameters, the zero point frequency shift of the original Fano resonance is calculated to be $\sim0.008\kappa_1$ and the width increase to be negligible. In Fig.~\ref{Fig6}a, the new Fano resonance emerges around $\omega_p-\omega_1+\Omega_1 \simeq -0.06\kappa_1$ which matches our calculation.

In Fig.~\ref{Fig6}b, we also plot the Stokes field in cavity $1$. It is interesting that a narrow dip is created inside the original single-peak lineshape when cavity $2$ is coupled to the OMS. The widths of the broad lineshape and the narrow dip are close to the widths of the original and newly-emerged Fano resonances of the antiStokes field in cavity $1$, respectively. The dip is caused by cavity $2$ adding an extra damping mechanism to the mechanical resonator and destructively interfering with the mechanical damping, so that it prevents the mechanical mode from aiding the generation of the Stokes field in cavity $1$.

\section{Fano resonances in quantum fluctuations of fields} \label{quantum}
\begin{figure}[phtb]
 \includegraphics[width=0.35\textwidth]{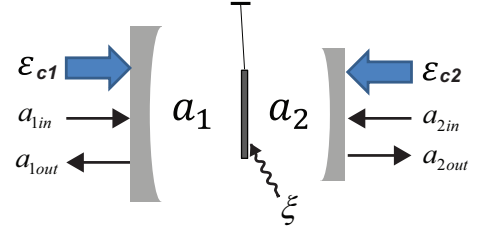}
 \caption{\label{Fig7}{(Color online) Schematic double-cavity OMS. Here $\mathcal{E}_{ci}$'s are coherent fields and $a_{i\text{in}}$'s are the quantum vacuum fields. $\xi$ is the Brownian noise.}}
\end{figure}

In the previous sections, we studied the OMS when the optical cavity is fed by both a detuned coupling field and a weak probe field, and found its output exhibits Fano resonance. Now we will study the quantum fluctuation of the cavity field without any input probe field, as illustrated in Fig.~\ref{Fig7}.
The quantum fluctuation of the cavity fields arises (i) direct from the fluctuation of the vacuum input and (ii) from the process of photon creation via oscillating mirror subjected to thermal noise. Those two mechanisms can interfere destructively creating a zero amplitude in the fluctuations of the cavity field.

In a double-cavity OMS, the quantum Langevin equations governing the operators $Q$, $P$, $a_i$ and $a_i^\dag$ are given by
\begin{equation}\label{40}
\begin{aligned}
		\dot{Q} & =\omega_{m}P, \\
		\dot{P} & =(g_{1}a_{1}^{\dagger}a_{1}-g_{2}a_{2}^{\dagger}a_{2})-\omega_{m}Q-\gamma_{m}P + \xi, \\
		\dot{a}_{1} & = -i(\omega_1-\omega_{c1}-g_1Q)a_{1} -\kappa_{1}a_{1} +\mathcal{E}_{c1} + \sqrt{2\kappa_1} a_\text{1in}, 	\\
		\dot{a}_{2} & = -i(\omega_2-\omega_{c2}+g_2Q)a_{2}-\kappa_{2}a_{2}+\mathcal{E}_{c2} + \sqrt{2\kappa_2} a_\text{2in}.
\end{aligned}
\end{equation}
where $a_\text{1in}$ and $a^\dag_\text{1in}$ are the input noise from cavity $1$ with correlation function $\langle a_\text{1in}(t)a^\dag_\text{1in}(t')\rangle =\delta(t-t')$ and $\xi$ stems form the thermal noise of the mechanical resonator at finite temperature with correlation function in frequency domain
\begin{equation}\label{4ins}
\langle\xi(\omega)\xi(\Omega)\rangle = 2\pi \frac{\gamma_{m}}{\omega_{m}}\omega\left[1+\coth\left(\frac{\hbar\omega}{2k_B T}\right)\right]\delta(\omega+\Omega),
\end{equation}
where $k_B$ is the Boltzmann constant and $T$ is the temperature of the environment of the mirror.

Eqs.~(\ref{40}) are difficult to solve because they are nonlinear. However, considering that the quantum fluctuation values around their steady states are relatively small, we can adopt the standard linearization method by separating the fluctuations from their mean values,
\begin{equation}\label{41}
    Q=Q_0 + \delta Q, \quad P=P_0 + \delta P, \quad a_i = a_{i0} + \delta a_i,
\end{equation}
for $i=1,2$. When expanding the products of two operators $A$ and $B$, we can make the approximation $\delta(AB)\approx A_0\delta B + B_0\delta A$ so that quantum Langevin equations are modified as
\begin{equation}\label{42}
    \begin{aligned}
        \delta\dot{Q} & =\omega_{m}\delta P, \\
		\delta\dot{P} & =g_{1}(a_{10}^*\delta a_1+a_{10}\delta a_1^\dag)-g_{2}(a_{20}^*\delta a_2+a_{20}\delta a_2^\dag) \\
            & \qquad -\omega_{m}\delta Q-\gamma_{m}\delta P + \xi, \\
		\delta\dot{a}_{1} & = -(\kappa_1 + i\Delta_1)\delta a_1 + ig_1a_{10}\delta Q + \sqrt{2\kappa_1} a_\text{1in}, 	\\
		\delta\dot{a}_{2} & = -(\kappa_2 + i\Delta_2)\delta a_1 - ig_2a_{20}\delta Q + \sqrt{2\kappa_2} a_\text{2in},
    \end{aligned}
\end{equation}
The coupling fields are absorbed in the steady state mean values, so they do not show explicitly in Eq.~(\ref{42}). The $a_{i0}$'s and $\Delta_i$'s are defined identically to Sec.~\ref{model}. In order to get the spectra of the fluctuations in the quantities $\delta Q$, $\delta P$, $\delta a_i$ and $\delta a_i^\dag$, we Fourier transform them into the frequency domain using $f(t)=\frac{1}{2\pi}\int_{-\infty}^{+\infty}f(\omega)\mathrm{e}^{-i\omega t}\mathrm{d}\omega$. By solving them, we obtain the fields in cavity $1$ containing the signature of the quantum fluctuations
\begin{align}\label{43}
    &\sqrt{2\kappa_1} \delta a_1(\omega) = E_1(\omega)a_\text{1in}(\omega) + F_1(\omega)a^\dag_\text{1in}(-\omega) \nonumber \\
    & \quad + E_2(\omega)a_\text{2in}(\omega) + F_2(\omega)a^\dag_\text{2in}(-\omega) + V(\omega)\xi(\omega),
\end{align}
and fluctuations $a_2(\omega)$ in cavity $2$ can be calculated similarly using the symmetry property of the double-cavity configuration. Note that the term $E_1(\omega)$ physically means that a noise photon at $\omega+\omega_{c1}$ produces a photon at frequency $\omega+\omega_{c1}$ where as the term $F_1(\omega)$ corresponds to the four wave mixing process where a photon of frequency $\omega_{c1}-\omega$ produces a photon of frequency $\omega_{c1}+\omega$. Similar interpretations apply to $E_2(\omega)$ and $F_2(\omega)$. Thus $E_1(\omega)$ and $F_1(-\omega)$ would have direct relation to the antiStokes and Stokes fields discussed in the earlier sections.  The mechanical noise can be suppressed by cooling down the environment temperature, though it is the dominant contribution to the fluctuations at high temperatures, and hence we omit the $\xi(\omega)$ term. Here $E_i(\omega)$'s, and $F_i(\omega)$'s are the functions given by
\begin{equation}\label{44}
\begin{aligned}
    E_1(\omega) &= -\frac{2\kappa_1}{D(\omega)} \frac{2iG_1^2}{(\kappa_1+i\Delta_1-i\omega)^2} + \frac{2\kappa_1}{\kappa_1+i\Delta_1-i\omega}, \\
    F_1(\omega) &= -\frac{2\kappa_1}{D(\omega)} \frac{2iG_1^2}{(\kappa_1-i\omega)^2+\Delta_1^2}, \\
    E_2(\omega) &= \frac{\sqrt{2\kappa_12\kappa_2}}{D(\omega)} \frac{2iG_1G_2}{(\kappa_1+i\Delta_1-i\omega)(\kappa_2+i\Delta_2-i\omega)}, \\
    F_2(\omega) &= \frac{\sqrt{2\kappa_12\kappa_2}}{D(\omega)} \frac{2iG_1G_2}{(\kappa_1+i\Delta_1-i\omega)(\kappa_2-i\Delta_2-i\omega)}, \\
    D(\omega) &= \sum_{i=1,2} \frac{4\Delta_iG_i^2}{(\kappa_i-i\omega)^2 + \Delta_i^2} - \frac{\omega_m^2-\omega^2-i\omega\gamma_m}{\omega_m}.
\end{aligned}
\end{equation}

The quadratures of the field in cavity $1$, which can be measured using homodyne detection, have the spectra defined as $\langle X_1(\Omega)X_1(\omega)\rangle = 2\pi S_{1X}\delta(\omega+\Omega)$ and $\langle Y_1(\Omega)Y_1(\omega)\rangle = 2\pi S_{1Y}\delta(\omega+\Omega)$ with $X_1=(a_1^\dag+a_1)/\sqrt{2}$ and $Y_1=i(a_1^\dag-a_1)/\sqrt{2}$. Now we calculate the fluctuation spectrum in the $X$ quadrature as
\begin{align}\label{45}
    &2\kappa_1 S_{1X}(\omega) \nonumber \\
    &= \frac12|E_1^*(-\omega)+F_1(\omega)|^2 + \frac12|E_2^*(-\omega)+F_2(\omega)|^2 \nonumber\\
    &=\frac12 \left|\frac{2\kappa_1}{\kappa_1+i\Delta_1+i\omega}\right|^2 \left|1-\frac{1}{D(\omega)}\frac{4\Delta_1G_1^2}{(\kappa_1+i\omega)^2+\Delta_1^2}\right|^2 \nonumber\\
    &\quad +\frac12 \left|\frac{\sqrt{2\kappa_12\kappa_2}}{\kappa_2+i\Delta_2+i\omega}\right|^2 \left|\frac{1}{D(\omega)}\frac{4\Delta_1G_1G_2}{(\kappa_1+i\omega)^2+\Delta_1^2}\right|^2.
\end{align}
We first study single-cavity OMS when $G_2=0$. The cavity field fluctuation in $X$ quadrature is given as
\begin{equation}\label{46}
    2\kappa_1 S_{1X}(\omega) \approx \frac{\kappa_1^2}{2\omega_m^2} \frac{(\omega-\omega_m)^2}{ (\omega-\omega_m-\frac{\Omega_1G_1^2}{\Omega_1^2+\kappa_1^2})^2 + (\frac{\kappa_1G_1^2}{\Omega_1^2+\kappa_1^2})^2}.
\end{equation}
We do not show the expressions of fluctuations $S_{1Y}(\omega)$ or $S_{1a}(\omega)$ since they do not exhibit Fano minimum. Equation~(\ref{46}) indicates a Fano lineshape, which has a minimum at $\omega=\omega_m$ and a maximum at $\omega=\omega_m + \frac{\Omega_1G_1^2}{\Omega_1^2+\kappa_1^2}$ with width $\Gamma_\text{qu}=\frac{\kappa_1G_1^2}{\Omega_1^2+\kappa_1^2}$ and asymmetry parameter $q=-\Omega_1/\kappa_1$. To see the Fano resonance, it is important to have $\kappa_1\gg\gamma_m$ and $G_i^2\gg\kappa_i\gamma_m$. The spectra $S_\text{1Xout}(\omega)$ and $S_\text{1Yout}(\omega)$ of the output field are different from the cavity fields by an amount of $a_\text{1in}$ using the input-output relation $a_\text{1out}=\sqrt{2\kappa_1}\delta a_1 - a_\text{1in}$.

\begin{figure}[phtb]
 \includegraphics[width=0.48\textwidth]{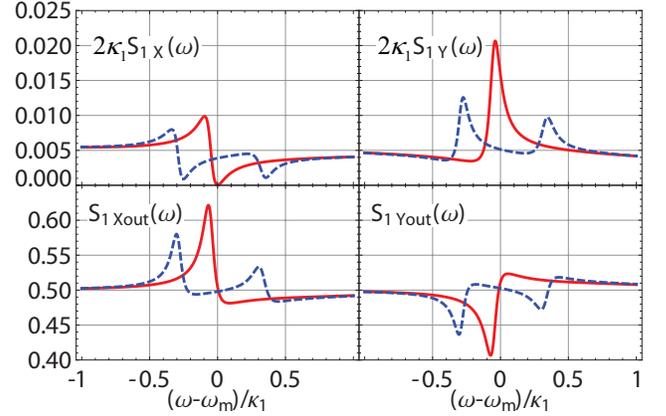}
 \caption{\label{Fig8}{(Color online) The spectra of the quadratures for the cavity fields and output fields for both single-cavity OMS (solid curves) and double-cavity OMS (dashed curves). The parameters used are the same to Fig.~\ref{Fig3}. }}
\end{figure}

We illustrate the spectra of the quadrature $S_\text{1X}(\omega)$, $S_\text{1Y}(\omega)$, $S_\text{1Xout}(\omega)$, and $S_\text{1Yout}(\omega)$ for both single-cavity OMS (solid curves) and double-cavity OMS (dashed curves) in Fig.~\ref{Fig8} using parameters as in Fig.~\ref{Fig5}.
From the solid curves in the figure, we see that the $S_\text{1X}(\omega)$ quadrature exhibits a clear Fano resonance. The Fano resonance has a zero point at frequency $\omega-\omega_m=0$ and has width $\Gamma_\text{qu}=0.1\kappa_1$, both of which match our calculation. The spectrum of the quadratures $S_\text{1Xout}(\omega)$ and $S_\text{1Yout}(\omega)$ of the output field also have typically asymmetric line shapes which are signatures of interferences. These spectra have similarities to the spectra for the Stokes field~(Fig.~\ref{Fig4}b). Note that the formula like Eq.~(\ref{46}) shows that the quadrature spectra are determined by the interference of the Stokes and antiStokes terms. The reason is that in the region of interest in the spectrum $S_\text{1X}(\omega)$, the term $E_1^*(-\omega)$ is approximately flat. When the second cavity is coupled to the system, we expect a splitting of the Fano lineshape in the spectrum of fluctuations following the the classical analysis of Sec.~\ref{double}. The splittings of the resonances separated by $0.6\kappa_1=2G_2$ appear in the dashed curves in Fig.~\ref{Fig8}. The splittings are due to the enhanced coupling strength by the increasing photon number in the cavities, which induced normal mode splitting of the cavity states.

\section{Conclusion}
In conclusion, we have shown how the asymmetric Fano line shapes can arise in optomechanics. We identify interfering pathways leading to the Fano resonances. In contrast to atomic systems, the coupling field can be used to tune Fano resonances using both the frequency and the power of coupling field. In fact, as displayed in Fig.~\ref{Fig1}b the coupling field opens up another coherent path way. We give explicit expressions for the width and the asymmetry parameter. The Fano resonances can be studied both via pump probe experiment and via the study of the quantum fluctuations in the output fields. The Fano minima are much more pronounced in the results of the pump probe experiments. The double cavity OMS produce double Fano minima .

\end{document}